\documentclass[aip, amsmath,amssymb, reprint, superscriptaddress, citeautoscript, ]{revtex4-1}
\usepackage{natbib}
\usepackage[version=4]{mhchem}
\usepackage{amsmath, amssymb}  
\usepackage{graphicx}    
\usepackage{lipsum}
\usepackage{color}  
\usepackage[dvipsnames]{xcolor}

\begin{document}

\title{Continuous Representation Theory of the Hierarchical Equations of Motion}

\author{Tianchu Li}
\affiliation{Department of Chemistry, University of Colorado Boulder, Boulder, Colorado 80309, USA\looseness=-1}
\author{Pranay Venkatesh}
\affiliation{Department of Chemistry, University of Colorado Boulder, Boulder, Colorado 80309, USA\looseness=-1}
\author{Andr\'es Montoya-Castillo}
\email{Andres.MontoyaCastillo@colorado.edu}
\affiliation{Department of Chemistry, University of Colorado Boulder, Boulder, Colorado 80309, USA\looseness=-1}

\begin{abstract}
The numerical stability of the hierarchical equations of motion (HEOM) depends on the representation of their Liouvillian. Here, we introduce a continuous gauge freedom through a generalized Bogoliubov transformation with the parameter $r$. It subsumes earlier phase space stabilization methods as special cases and explains both their stabilizing mechanism and their limitations. We find that gauge optimization balances the suppression of hierarchy growth against errors in reconstructing the reduced density operator, and show that the factor $e^{2r}-1$ controls the feedback strength, with earlier methods located at the boundary for reconstruction-error amplification, $r=\frac{1}{2}\ln 2$. For spin–boson models, this boundary gauge recovers known stabilizations for undamped and weakly damped modes. In the weakly damped case, an interior gauge further reduces the trace error by more than an order of magnitude. For the Holstein model, interior gauges reduce trace drift by about an order of magnitude relative to the previously adopted fixed representation at the same bond dimension. This improvement stabilizes error-sensitive calculations of equilibrium autocorrelation functions. Our formulation thus extends fixed stabilization methods to a continuous family of HEOM representations generated by linear canonical transformations, and establishes a framework for improving numerical stability and accuracy through continuous gauge selection.
\end{abstract}

\maketitle

\begin{figure}[b]
\centering
\includegraphics[width=6.40cm]{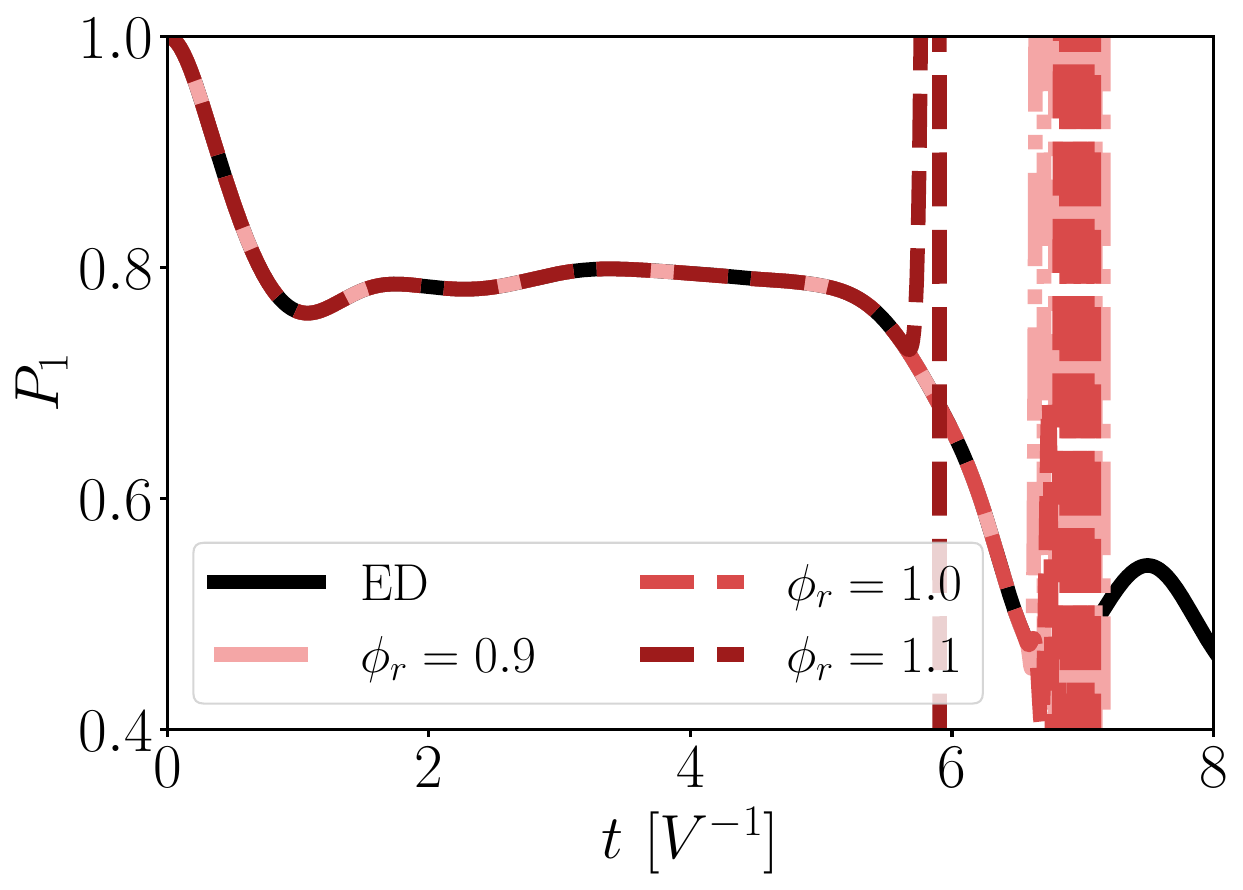}
\vspace{-10pt}
\caption{
Divergence of conventional and $\mathbf{M}$-transformed HEOM dynamics for the spin--boson model under strong coupling to a single undamped mode ($c=2\sqrt{2}$, $\omega=1.0$, $\beta=1.0$). Population of state $|1\rangle$ obtained using exact diagonalization (benchmark), conventional HEOM ($\phi_r=1.0$), and $\mathbf{M}$-transformed HEOM with $\phi_r=0.9$ and $1.1$. The gauge parameter $\phi_r$ is defined in Ref.~\onlinecite{li2026hidden}.
}
\vspace{-10pt}
\label{fig1}
\end{figure}

\textit{Introduction} --- The hierarchical equations of motion (HEOM) provide a nonperturbative and non-Markovian framework for open quantum dynamics~\cite{tanimura89,tanimura06,tanimura20,yan14a,bai24,xu2026colloquium}. Despite its broad adoption, HEOM can exhibit severe numerical instabilities that depend sensitively on the structure of the hierarchy Liouvillian~\cite{yan20,krug2023stability,li22,dunn19,ikeda20}. We have recently uncovered a continuous gauge freedom in complex-pole HEOM that remains analytically exact while reshaping the hierarchy Liouvillian to numerically suppress instabilities associated with positive-real-part eigenvalues and non-normal error amplification~\cite{li2026hidden}. Unlike projection-based remedies that remove unstable modes from the truncated hierarchy~\cite{dunn19}, gauge transformations leave the exact hierarchy intact and change only how one represents it. Despite this advantage, our previously proposed gauge transformation does not eliminate numerical instabilities in the undamped, single-mode environment~\cite{liu14}. In this case, all gauge choices examined in our earlier work eventually yielded divergent dynamics at finite hierarchy truncation, with the gauge that recovered conventional HEOM remaining stable the longest (see Fig.~\ref{fig1}). Thus, while optimizing within this gauge family could tame instabilities over the under- to over-damped limits, it offered no advantage in the fully undamped limit. Crucially, because an undamped single mode represents the zero-linewidth limit of a narrow spectral peak, similar limitations may be expected to arise for environments with sufficiently narrow but finitely broadened spectral peaks common across the chemistry and physics of charge, energy, and information transport. Indeed, such narrow peaks routinely arise from weakly damped intramolecular vibrations coupled to excitation energy transfer in photosynthetic complexes and electron transfer in cryptochromes~\cite{firmino16, novoderezhkin2017exciton, gustin2023mapping, tanaka09, gelzinis2017spectroscopic, korol2025high, fuller14, romero14}. This limitation thus motivates identifying more general gauge families that remain stable across the entire physical range, from the undamped to overdamped limits.

Here, we introduce a larger family of HEOM gauges that unifies earlier phase space stabilization methods as special cases and reveals how to generalize their applicability.
Building on the phase space reformulations introduced in Refs.~\onlinecite{shi09c, li25, xing22a, xu2023universal, xu2025quantum,chen2024bexcitonics}, we construct a continuous family of hierarchy representations through a generalized Bogoliubov transformation~\cite{balian1969nonunitary,yuen1976two} and show that earlier phase space-based stabilization methods~\cite{li22,yan20} become specific choices within this family. Unlike our previous phase ($\mathbf{M}$) gauge that mixes auxiliary creation operators and annihilation operators only within their respective sets, our current transformation mixes the two sets, achieving greater generality. Through analytical and numerical analyses, we show when a previously proposed stabilization works, why it incurs trace drift, and how moving continuously away from it within our gauge family recovers its stability while reducing that error, from the undamped limit to finitely broadened peaks. We further demonstrate the applicability of our framework to other physical systems, including the Holstein model of small polaron formation~\cite{holstein59,holstein59a,mendes2026coherent,bhattacharyya2025nonequilibrium,troisi07} applicable to organic semiconductors, transition metal oxides, and other semiconducting materials.

\textit{Method} --- To establish this framework, we consider the Caldeira--Leggett model~\cite{caldeira83,caldeira83a,weiss12}, a standard starting point for developing HEOM, in which a system operator $f(q)$ is linearly coupled to a harmonic bath,
\begin{align}
    H = H_{\rm sys}(p,q) + \frac{1}{2}\sum_j \left[p_j^2+\omega_j^2\left(x_j-\frac{c_jf(q)}{\omega_j^2}\right)^2\right].
    \label{eq:hcl}
\end{align}
Here, $H_{\rm sys}(p,q)$ is the system Hamiltonian, $x_j$ and $p_j$ are the mass-weighted coordinate and momentum, and $\omega_j$ and $c_j$ are the frequency and coupling coefficient of the $j$th bath oscillator. We set $\hbar=1$ throughout.

To expose the structure of the additional gauge freedom, we first consider an undamped single-mode environment with frequency $\omega$ and coupling coefficient $c$. For a thermal bath at inverse temperature $\beta$, the correlation function $C(t)=\langle F(t)F(0)\rangle_{\beta}$, with $F=cx$, is
\begin{align}
    C(t) = &\frac{c^2}{4\omega}\left[\coth\left(\frac{\beta\omega}{2}\right)-1\right]e^{i\omega t}\nonumber \\
    &+\frac{c^2}{4\omega}\left[\coth\left(\frac{\beta\omega}{2}\right)+1\right]e^{-i\omega t}.
\end{align}
This two-exponential form yields the following HEOM:
\begin{align}
    &\dot\rho_{m,n} = -i\left[\mathcal L_s - \omega(m-n)\right]\rho_{m,n}\nonumber \\
    &\ \  -i\left[f(q), \alpha^{(1)}\sqrt{m+1}\rho_{m+1,n} + \alpha^{(1)}\sqrt{n+1}\rho_{m,n+1}\right]\nonumber \\
    &\ \  -i\left[f(q),  \alpha^{(1)}\sqrt{m}\rho_{m-1,n} + \alpha^{(1)}\sqrt{n}\rho_{m,n-1}\right]\nonumber \\
    &\ \ +i\left\{f(q),\alpha^{(2)}\sqrt{m}\rho_{m-1,n} - \alpha^{(2)}\sqrt{n}\rho_{m,n-1}\right\},
    \label{eq:sheom}
\end{align}
where $\alpha^{(1)} = c\sqrt{\coth(\beta\omega/2)/4\omega}$, $\alpha^{(2)} = c\sqrt{\tanh(\beta\omega/2)/4\omega}$. This single-mode environment corresponds to the undamped limit of complex-pole HEOM\cite{liu14,tanaka09,xu22}, in which the poles $z_{\pm}=\pm i\omega$ have vanishing real parts, yielding a purely oscillatory bath correlation function without exponential decay. 

To render the representation freedom visible, we recast this hierarchy in phase space using auxiliary ladder operators that raise and lower the indices $m$ and $n$. These operators satisfy canonical commutation relations but are not required to be Hermitian conjugates of one another. The phase space equation of motion can be written as,
\begin{align}
    \dot \rho(x,p) =& -i\left(\mathcal L_{s} + \omega \tilde c_1^+\tilde c_1 -\omega \tilde c_2^+\tilde c_2\right)\rho(x,p) \nonumber \\
    & -i\alpha^{(1)} (\tilde c_1 + \tilde c_2) [f(q),\rho(x,p)] \nonumber \\
    & -i\alpha^{(1)} (\tilde c_1^+ + \tilde c_2^+) [f(q),\rho(x,p)] \nonumber \\
    & +i\alpha^{(2)} (\tilde c_1^+ - \tilde c_2^+) \{f(q),\rho(x,p)\}.
    \label{eq:mqcl}
\end{align}
Here, $\tilde c_i=e^{-S}c_i e^{S}$, with $S=(x^2+p^2)/4$, and $c_i$ are the ladder operators defined in Ref.~\onlinecite{li2026hidden}. In the single-mode limit, the phase space variables $x$ and $p$ coincide with the coordinate and momentum of the bath oscillator appearing in the Hamiltonian. We expand the auxiliary phase space distribution as
\begin{align}\label{eq:rdm-expansion}
    \rho(x,p;t) = \sum_{m,n}\frac{1}{\sqrt{m!n!}}\rho_{m,n}(t)\phi_{m,n}(x,p),
\end{align}
using the Gaussian basis
\begin{align}
    \phi_{m,n}(x,p) = \frac{1}{2\pi}(\tilde c_1^+)^m(\tilde c_2^+)^n|0,0\rangle,
\end{align}
where $|0,0\rangle=e^{-2S}$ is the reference state. Substituting this expansion into the phase space equation of motion in Eq.~\eqref{eq:mqcl} recovers Eq.~\eqref{eq:sheom}, with the reduced density operator given by
\begin{align}
    \rho_s(t)=\int \mathrm{d}x\,\mathrm{d}p\,\rho(x,p;t)=\rho_{0,0}(t).
\end{align}

Invoking the transformation matrix introduced as the phase gauge in Ref.~\onlinecite{li2026hidden},
\begin{align}
    \mathbf M(\theta)=
    \frac{1}{\sqrt{2\sin\theta}}\begin{pmatrix}
        -e^{-i\theta} & 1\\
        -e^{i\theta} & 1
    \end{pmatrix},
\end{align}
we apply $\mathbf M^{-1}(\pi/2)$ to transform the auxiliary raising and lowering operators separately, $(\tilde c_1^+,\tilde c_2^+) \mapsto (\tilde a^+,\tilde b^+)$ and $(\tilde c_1,\tilde c_2) \mapsto (\tilde a,\tilde b)$, preserving the canonical commutation relations. This yields
\begin{align}
    \dot{\rho}(x,p) ={}& -\left(i\mathcal L_s+\omega\tilde a^+\tilde b-\omega\tilde a\tilde b^+\right)\rho(x,p)\nonumber\\
    &-i\sqrt{2}\alpha^{(1)}(\tilde b+\tilde b^+)\left[f(q),\rho(x,p)\right]\nonumber\\
    &-\sqrt{2}\alpha^{(2)}\tilde a^+\left\{f(q),\rho(x,p)\right\},
    \label{eq:mqcl1}
\end{align}
where 
\begin{subequations} 
\begin{align} 
    \tilde a&=\frac{\partial}{\partial x}+x,\quad&\tilde b&=\frac{\partial}{\partial p}+p,\\ 
    \tilde a^+&=-\frac{\partial}{\partial x},\quad&\tilde b^+&=-\frac{\partial}{\partial p}. 
\end{align} 
\end{subequations} 
This equation is equivalent to the frictionless limit of the quantum Fokker--Planck equation (QFPE), or the mixed quantum--classical Liouville equation (MQCL)~\cite{kapral99,garg85,kapral06,ando03}. 

The ladder-operator representation above is not unique. While the transformation ${\mathbf M}^{-1}(\pi/2)$ mixes raising operators and lowering operators separately, a further freedom arises from mixing the two types while preserving their canonical commutation relations. To interrogate the potential benefits of this freedom, we introduce a one-parameter generalized Bogoliubov transformation,
\begin{align}
    \begin{pmatrix}
        \bar \xi\\
        \bar \xi^+
    \end{pmatrix}
    =
    \mathbf {B}(r)
    \begin{pmatrix}
        \tilde \xi\\
        \tilde \xi^+
    \end{pmatrix},
    \qquad
    \mathbf {B}(r)=
    \begin{pmatrix}
        e^{-r} & 0\\
        2\sinh r & e^r
    \end{pmatrix},
\end{align}
where $r$ is a real gauge parameter and $\xi=a,b,c$. This transformation preserves the canonical commutation relation $[\bar c,\bar c^+]=1$ and obeys the composition law
\begin{align}
    \mathbf {B}(r_1)\mathbf {B}(r_2)=\mathbf {B}(r_1+r_2).
\end{align}
Thus, $\mathbf {B}(r)$ forms a one-parameter group, with $\mathbf {B}(0)=\mathbf I$ and $\mathbf {B}^{-1}(r)=\mathbf {B}(-r)$. 

Applying $\mathbf {B}(r)$ to Eq.~\eqref{eq:mqcl1} and subsequently transforming the resulting ladder operators with $\mathbf M(\pi/2)$ yields a new phase space equation. Expanding the phase space distribution in the basis,
\begin{align}
    \phi_{m,n}(x,p) = \frac{1}{2\pi}(\bar c_1^+)^m(\bar c_2^+)^n|0,0\rangle,
\end{align}
we obtain the following $r$-dependent HEOM:
\begin{widetext}
\begin{align}
    \dot\rho_{m,n} = &-i\left[\mathcal L_s - \omega(m-n)\right]\rho_{m,n}-i\alpha^{(1)}e^{r}\left[f(q),\sqrt{m+1}\rho_{m+1,n} + \sqrt{n+1}\rho_{m,n+1}\right] \nonumber \\
    &-i\alpha^{(1)}e^{-r}[f(q),\sqrt{m}\rho_{m-1,n}] + i\alpha^{(2)}e^{-r}\{f(q),\sqrt{m}\rho_{m-1,n}\}\nonumber \\
    &-i\alpha^{(1)}e^{-r}[f(q),\sqrt{n}\rho_{m,n-1}] - i\alpha^{(2)}e^{-r}\{f(q),\sqrt{n}\rho_{m,n-1}\}\nonumber \\
    &+2i\alpha^{(1)}\sinh r[f(q),\sqrt{m+1}\rho_{m+1,n}] + 2i\alpha^{(2)}\sinh r\{f(q),\sqrt{m+1}\rho_{m+1,n}\}\nonumber \\
    &+2i\alpha^{(1)}\sinh r[f(q),\sqrt{n+1}\rho_{m,n+1}] - 2i\alpha^{(2)}\sinh r\{f(q),\sqrt{n+1}\rho_{m,n+1}\}.
    \label{eq:sheom1}
\end{align}
\end{widetext}
In this transformed HEOM, one can recover the reduced density operator as follows
\begin{align}
    \rho_s(t)=\int \mathrm{d}x\,\mathrm{d}p\,\rho(x,p;t)=\sum_{k=0}^{\infty}(e^{2r}-1)^k\rho_{k,k}(t).
    \label{eq:recons}
\end{align}
Unlike conventional HEOM and the phase gauge HEOM of Ref.~\onlinecite{li2026hidden}, where $\rho_s(t)=\rho_{0,0}(t)$, \textit{this representation reconstructs the reduced density operator as a weighted sum of the zeroth-order ADO and higher-order ADOs with equal indices}. One recovers the conventional extraction rule at $r=0$. 

Our present formulation brings previous attempts to tame HEOM instabilities under a common umbrella while offering a sufficiently flexible and principled path to identify the transformation that optimizes stability across physical limits encoded in the shapes and magnitudes of spectral densities. For instance, one can recover the earlier proposal to stabilize single-mode HEOM\cite{yan20} as a special case of our transformation where $r=\frac{1}{2}\ln 2$. At this value, the transformed ladder operators take the standard harmonic-oscillator form, $\bar a=(x+\partial_x)/\sqrt{2}$ and $\bar a^+=(x-\partial_x)/\sqrt{2}$~\cite{mahan13}, with analogous expressions for $\bar b$ and $\bar b^+$ obtained by replacing $x$ with $p$\footnote{Note that $x$ and $p$ are not canonical conjugates, consistent with previous phase space reformulations~\cite{yan20,li22,miller01}.}. For Brownian oscillator (BO) spectral densities, our gauge transformation also recovers the stabilized HEOM formulation of Ref.~\onlinecite{li22} by first applying $\mathbf M^{-1}(\theta_k)$ and then $\mathbf {B}(\frac{1}{2}\ln 2)$. Similarly, one can understand the recent advance in Ref.~\onlinecite{gatto2026improving} as a particular choice of gauge in our representation theory: one first decomposes the complex exponential bath-correlation terms into damped sine and cosine components and then applies $\mathbf {B}(\frac{1}{2}\ln2)$ to the resulting hierarchy. Our framework further provides the theoretical infrastructure to perform a principled search over the family of gauge transformations that yield optimal stability. 

Beyond unifying these stabilization schemes, our framework recovers the established equivalence between HEOM and the pseudomode formulation at zero temperature as a particular gauge choice\cite{xu2023universal,muller2026one,pleasance2020generalized,park2024quasi,tamascelli2018nonperturbative}. Applying the composite transformation $\mathbf M(\pi/2)\mathbf B(\frac{1}{2}\ln 2)\mathbf M^{-1}(\pi/2)$ to each auxiliary-mode pair associated with $\bar c_1^+$ and $\bar c_2^+$ yields the pseudomode Liouvillian expressed in the Fock basis. The auxiliary multi-indices $\mathbf m$ and $\mathbf n$ then label the ket and bra occupations of the pseudomodes, respectively. The transformed ADOs correspond to the system-operator-valued matrix elements $\langle\mathbf m|\varrho(t)|\mathbf n\rangle$ of the joint system–pseudomode density operator. In this representation, the multimode generalization of Eq.~\eqref{eq:recons} at $r=\frac{1}{2}\ln 2$ becomes $\rho_s(t)=\sum_{\mathbf n}\rho_{\mathbf n,\mathbf n}(t)=\operatorname{Tr}_{\mathrm{PM}}\varrho(t)$, identifying the reconstruction of the reduced density operator with the partial trace over the pseudomodes.

More generally, compositions of the phase $\mathbf M$ and Bogoliubov $\mathbf B$ transformations generate a continuous family of equivalent hierarchy representations. We derive the resulting HEOM for these and their corresponding reduced density operator reconstruction in the Supporting Information (SI)~Sec.~I, along with their extension to modes with finite linewidths. One can optimize the parameters of the $\mathbf M$ transformation using the low-cost gauge optimization protocol introduced in Ref.~\onlinecite{li2026hidden}. We discuss the choice of $r$ in the $\mathbf {B}$ transformation below.

\begin{figure*}[t]
\centering
\includegraphics[width=\textwidth]{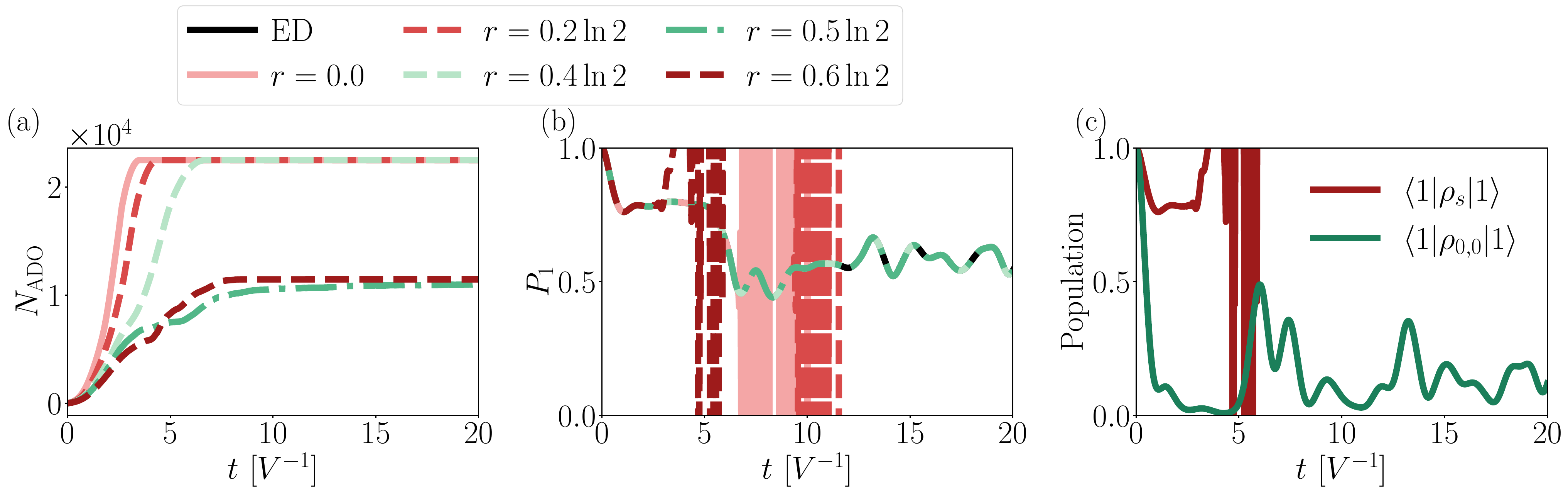}
\vspace{-20pt}
\caption{Performance of gauge-transformed HEOM for the spin--boson model under strong coupling to a single undamped mode ($c=2\sqrt{2}$, $\omega=1.0$, $\beta=1.0$), using the transformation sequence $\mathbf{M}\mathbf{B}(r)\mathbf{M}^{-1}$. (a) Number of active ADOs as a function of time for conventional HEOM ($r=0$) and gauge-transformed HEOM with $r=0.2\ln 2$, $r= 0.4\ln 2$, $r=0.5\ln 2$, and $r=0.6\ln 2$. (b) Population of state $|1\rangle$ obtained using the same HEOM calculations as in panel (a), with exact diagonalization as a benchmark. (c) Comparison of the $|1\rangle$ populations extracted from the reconstructed reduced density operator $\rho_s(t)$ and the zeroth-tier ADO $\langle 1|\rho_{0,0}(t)|1\rangle$ for $r=0.6\ln 2$. We use shades of red (green) to indicate gauge choices that yield divergent (well-behaved) population dynamics.
}
\label{fig2}
\vspace{-10pt}
\end{figure*}

\textit{Results and discussion} --- To assess the effect of our proposed gauge transformation, we illustrate our results on the spin--boson model, with the two-level system Hamiltonian
\begin{align}
    H_{\rm sys}=\frac{\epsilon}{2}\sigma_z+V\sigma_x,
    \label{eq:sb}
\end{align}
and the system--bath coupling operator $f(q)=\sigma_z$~\cite{nitzan06}. Here, $\epsilon$ is the energy bias, $V$ is the electronic coupling, and $\sigma_z$ and $\sigma_x$ are Pauli matrices. Working in reduced units of $V$, we set $\epsilon = 0$ and inverse temperature $\beta = V^{-1}$. The initial state is $\rho_{\mathrm{tot}}(0)=|1\rangle\langle 1|\otimes e^{-\beta h_b}/Z_b$, where $Z_b=\operatorname{Tr}_b(e^{-\beta h_b})$ is the bath partition function. We propagate the hierarchy using an on-the-fly filtering algorithm~\cite{shi09b}, which dynamically retains or discards auxiliary density operators (ADOs) according to their amplitudes.

We begin by examining how the gauge parameter $r$ can control the numerical stability of the transformed HEOM for an undamped single-mode environment. Figures~\ref{fig2}(a) and~\ref{fig2}(b) show that the number of ADOs in conventional HEOM grows rapidly, and the dynamics becomes unstable shortly after this number reaches the imposed upper bound. This observation suggests that the chosen hierarchy truncation is insufficient for the undamped single-mode dynamics. Without bath damping, higher-order ADOs are not directly suppressed by dissipative decay, allowing the hierarchy to continue growing during propagation. This underscores the need for a better HEOM generator that slows the growth in the number of ADOs while maintaining accuracy, stabilizing HEOM propagation.

To understand how the gauge transformation can suppress the growth in the number of active ADOs and improve the numerical stability of HEOM propagation, we analytically examine the couplings between adjacent hierarchy orders in Eq.~\eqref{eq:sheom1}. In HEOM, `$-$' terms, $\rho_{n^-}$, drive the evolution of higher-order ADOs, whereas `$+$' terms, $\rho_{n^+}$, provide feedback to lower orders. In Eq.~\eqref{eq:sheom1}, the coefficients of the `$+$' and `$-$' terms in the commutator are both $e^{-r}$. For positive $r$, this factor weakens both the driving toward higher orders and the feedback toward lower orders, potentially slowing the growth in the number of ADOs. Meanwhile, the anticommutator contributions involving `$-$' terms are also scaled by $e^{-r}$, whereas those involving `$+$' terms carry the factor $2\sinh r$. Increasing $r$ therefore weakens the driving toward higher orders but strengthens the anticommutator-mediated feedback toward lower orders. The ratio of these feedback and driving coefficients is $e^{2r}-1$, which remains at most unity for $0<r\leq\frac{1}{2}\ln 2$ and exceeds unity for $r>\frac{1}{2}\ln 2$. When this ratio remains below unity, the driving toward higher orders is suppressed while the feedback coefficient remains smaller than the driving coefficient, potentially limiting the growth in the number of ADOs. When this ratio exceeds unity, one might expect the weaker `$-$' terms to slow the growth in the number of ADOs at short times, but for the stronger feedback from higher orders to amplify lower-order amplitudes and subsequently activate additional higher-order ADOs at longer times, accelerating the growth of the hierarchy. We therefore conjecture that choosing $0<r\leq\frac{1}{2}\ln 2$ can more effectively suppress the long-time growth in the number of ADOs and therefore stabilize HEOM.

The results in Fig.~\ref{fig2}(b) support this interpretation. For the gauges tested in the range $0\leq r\leq \frac{1}{2}\ln 2$, increasing $r$ suppresses the growth in the number of active ADOs. The dynamics eventually diverges for $r=0.2\ln 2$, whereas no divergence arises for $r=0.4\ln 2$ and $0.5\ln 2$ over the simulated time. Increasing $r$ further from $0.5\ln 2$ to $0.6\ln 2$ initially slows ADO growth, but this trend reverses at longer times. This crossover reflects the competition between weaker driving of higher-order ADOs and stronger feedback to lower orders, and suggests that increasing $r$ beyond $\frac{1}{2}\ln 2$ does not improve stability. Under on-the-fly filtering, the number of active ADOs for both $r=0.5\ln 2$ and $r=0.6\ln 2$ eventually plateaus below the prescribed maximum ADO count over the simulated time.
\begin{figure}[b]
\vspace{-15pt}
\centering
\includegraphics[width=8.7cm]{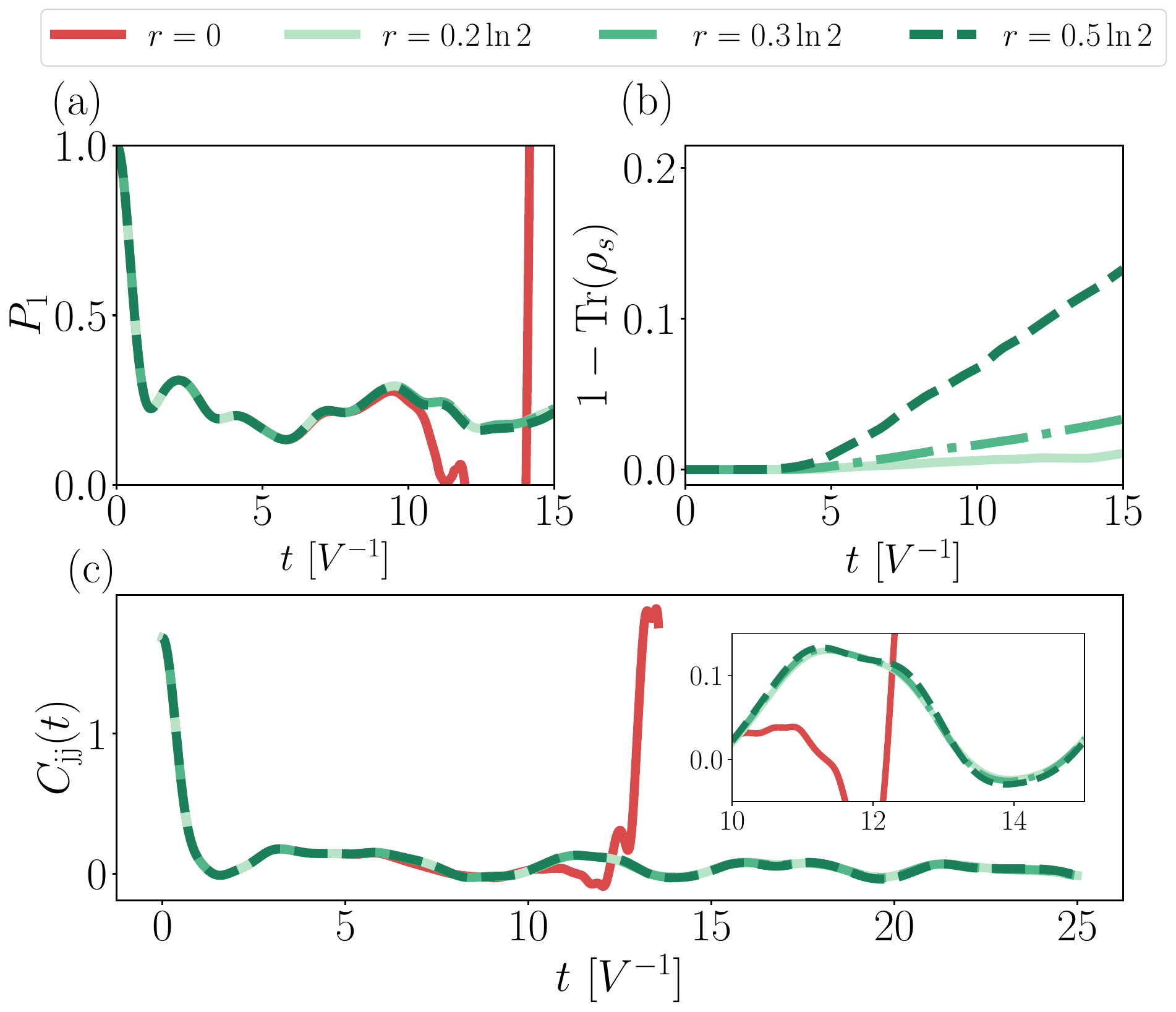}
\caption{Performance of our $\mathbf{B}$-transformed HEOM for a $5$-site Holstein model ($c=2\sqrt{2}$, $\omega=1.0$, $\beta=1.0$). (a) Population of the first site obtained using conventional HEOM ($r=0$), and $\mathbf{B}$-transformed HEOM with $r=0.2\ln 2$, $r=0.3\ln 2$, and $0.5\ln 2$. (b) Trace drift for the reconstructed reduced density operator, $1-\mathrm{Tr}\,\rho_s(t)$, for the three nonzero gauge values in panel (a). (c) Current autocorrelation function $C_{\rm jj}(t)$ for all the gauge values in panel (a).
}
\label{fig3}
\end{figure}

\begin{figure*}[t]
\centering
\includegraphics[width=\textwidth]{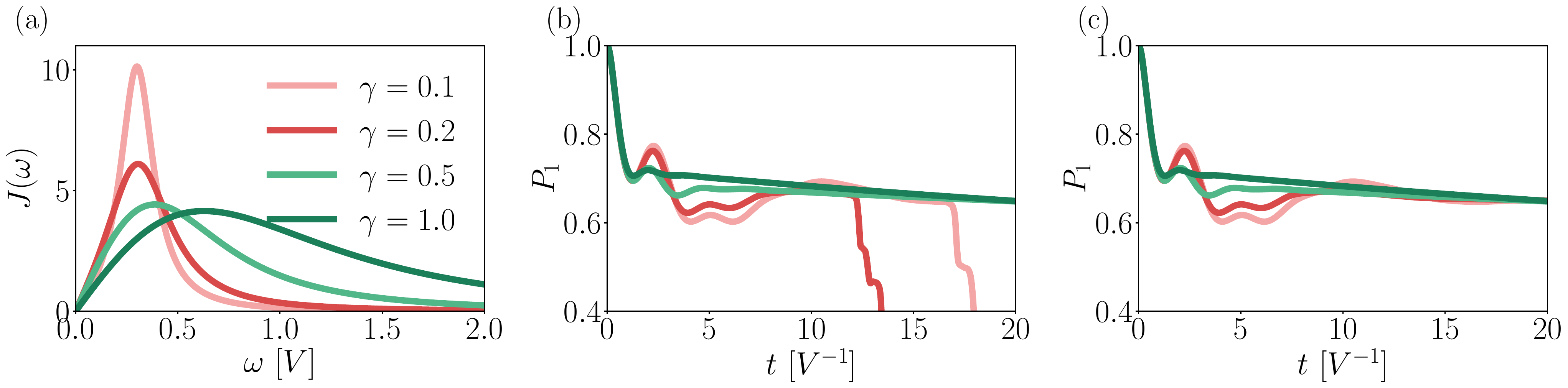}
\vspace{-10pt}
\caption{Performance of conventional HEOM versus a compound $\mathbf{B}$-$\mathbf{M}^{-1}$-transformed HEOM for underdamped spectral densities of finite width benchmarked on a spin--boson model at fixed reorganization energy $\lambda=3.125$ and at $\beta=1.0$, with $\omega_b=0.3$ and $\gamma=0.1$, $0.2$, $0.5$, and $1.0$. (a) Spectral densities. Population of state $|1\rangle$ obtained using (b) conventional HEOM and (c) transformed HEOM with the transformation sequence $\mathbf{B}(\frac{1}{2}\ln 2)\mathbf{M}^{-1}(\arctan(\omega_b/\gamma))$. 
}
\label{fig4}
\vspace{-10pt}
\end{figure*}
However, a plateau in the number of ADOs does not guarantee stable reduced dynamics. Figure~\ref{fig2}(c) shows the zeroth-order ADO $\rho_{0,0}(t)$ remains stable for $r=0.6\ln 2$, whereas its reconstructed reduced density operator $\rho_s(t)$ diverges. This behavior is consistent with the reconstruction weights $(e^{2r}-1)^k$, which grow exponentially with $k$ when $r>\frac{1}{2}\ln 2$ and can amplify errors in higher-order ADOs. For $0\leq r\leq\frac{1}{2}\ln 2$, these weights remain at most $1$, avoiding this source of error amplification. Together with the suppression of lower-order source terms for positive $r$, these results identify $0<r\leq\frac{1}{2}\ln 2$ as a favorable range for gauge selection. Within this interval, increasing $r$ can improve the numerical behavior of the ADOs $\rho_{m,n}$ but also increase the weights of higher-order ADOs in the reconstruction of $\rho_s$, potentially increasing the resulting error. For the undamped single-mode spin--boson model considered here, $r=\frac{1}{2}\ln 2$, which recovers the stabilization method of Ref.~\onlinecite{yan20}, provides the best balance between numerical stability and reconstruction accuracy. \textit{Gauge selection therefore involves a trade-off between improved hierarchy propagation and greater sensitivity of the reconstructed reduced density operator to higher-order ADO errors.}

Having identified the advantageous gauge range for the single-mode spin--boson model, which couples the system to a single mode antisymmetrically, we next examine whether the same gauge is optimal for systems coupled to multiple independent baths. To this end, we consider the Holstein model, which is widely used to study charge transport in organic molecular crystals and other materials with strong electron--phonon coupling~\cite{holstein59, holstein59a, li24, song15, xing22, bhattacharyya2025nonequilibrium, jankovic2025charge, jankovic2022spectral}. In this model, each electronic site couples to an independent local vibrational mode. In the dilute limit, the system Hamiltonian takes the form
\begin{align}
    H_{\rm sys}=\sum_{m=1}^{N}\Big[\epsilon_m |m\rangle\langle m|-V\left(|m\rangle\langle m+1|+\rm{H.c.}\right)\Big],
\end{align}
and the system--bath coupling operators $f_m=|m\rangle\langle m|$. Here, $\epsilon_m$ is the site energy, $V$ is the nearest-neighbor hopping amplitude. Each site couples to an independent local harmonic oscillator through the coordinate-space coupling defined in Eq.~\eqref{eq:hcl}. We consider a chain of $5$ sites with periodic boundary conditions with $\epsilon_m = 0, V=1$. The choice $r=\frac{1}{2}\ln 2$ recovers the previously developed stabilization method~\cite{yan20}, which exhibits unwanted trace drift in the reconstructed reduced density operator (see Fig.~\ref{fig3}(b)). Preserving the trace is important for reliable calculations of physical observables, especially fluctuations~\cite{bhattacharyya2024anomalous,song15,yang2025hierarchical,li24}, since normalization errors can affect expectation values, spectroscopic observables, and transport coefficients such as mobility and frequency-dependent conductance. We find that substantial trace drift persists in the previously developed stabilization method even when the matrix product state bond dimension is increased to $\chi=600$ (see Fig.~S1). 

Could varying $r$ in $(0,\frac{1}{2}\ln 2]$ reduce this drift while retaining improved numerical stability? Figure~\ref{fig3} compares the population dynamics at the first site obtained using conventional HEOM ($r=0$) and gauge-transformed HEOM with different values of $r$. All our Holstein simulations employ a matrix product state representation of the hierarchy~\cite{shi18, yan21a, ke22a, ke2023tree, borrelli21, chen2025tree, rodriguez2026tenso, guan24, li2026molecular} (see SI~Sec.~II for details). Conventional HEOM exhibits divergent dynamics, whereas all gauge-transformed HEOM calculations remain numerically stable over the simulated time. For $r=\frac{1}{2}\ln 2$, corresponding to the previously developed approach to stabilizing single-mode HEOM~\cite{yan20}, the reconstructed reduced density operator instead shows trace drift. This is consistent with accumulated errors from higher-order ADOs, which contribute to the reconstruction with unit weights. Among the gauge values tested, $r=0.2\ln 2$ performs best, yielding stable dynamics while maintaining the trace of the reconstructed reduced density operator closest to unity, with $1-\mathrm{Tr}\,\rho_s\approx 10^{-2}$ at $t=15$, compared with $1-\mathrm{Tr}\,\rho_s\approx 0.13$ for $r=\frac{1}{2}\ln 2$. To further illustrate the impact of trace drift on dynamical observables, Fig.~\ref{fig3}(c) shows the current autocorrelation function $C_{\rm jj}(t)$, which characterizes equilibrium current fluctuations. This correlation function encodes the frequency-resolved conductance and the terahertz response of materials\cite{bhattacharyya2024anomalous,bhattacharyya2024mori,fetherolf20}. The results converge for $r=0.2\ln 2$ and $r=0.3\ln 2$, but not for $r=0.5\ln 2$. We provide details of the equilibrium correlation function calculations using gauge-transformed HEOM in SI~Sec.~III. Importantly, our gauge-transformed HEOM also applies beyond the dilute limit, as demonstrated by stable dynamics for two spinless fermions on a four-site Holstein model at half filling (see SI~Sec.~IV). 

\begin{figure}[b]
\vspace{-10pt}
\centering
\includegraphics[width=7.00cm]{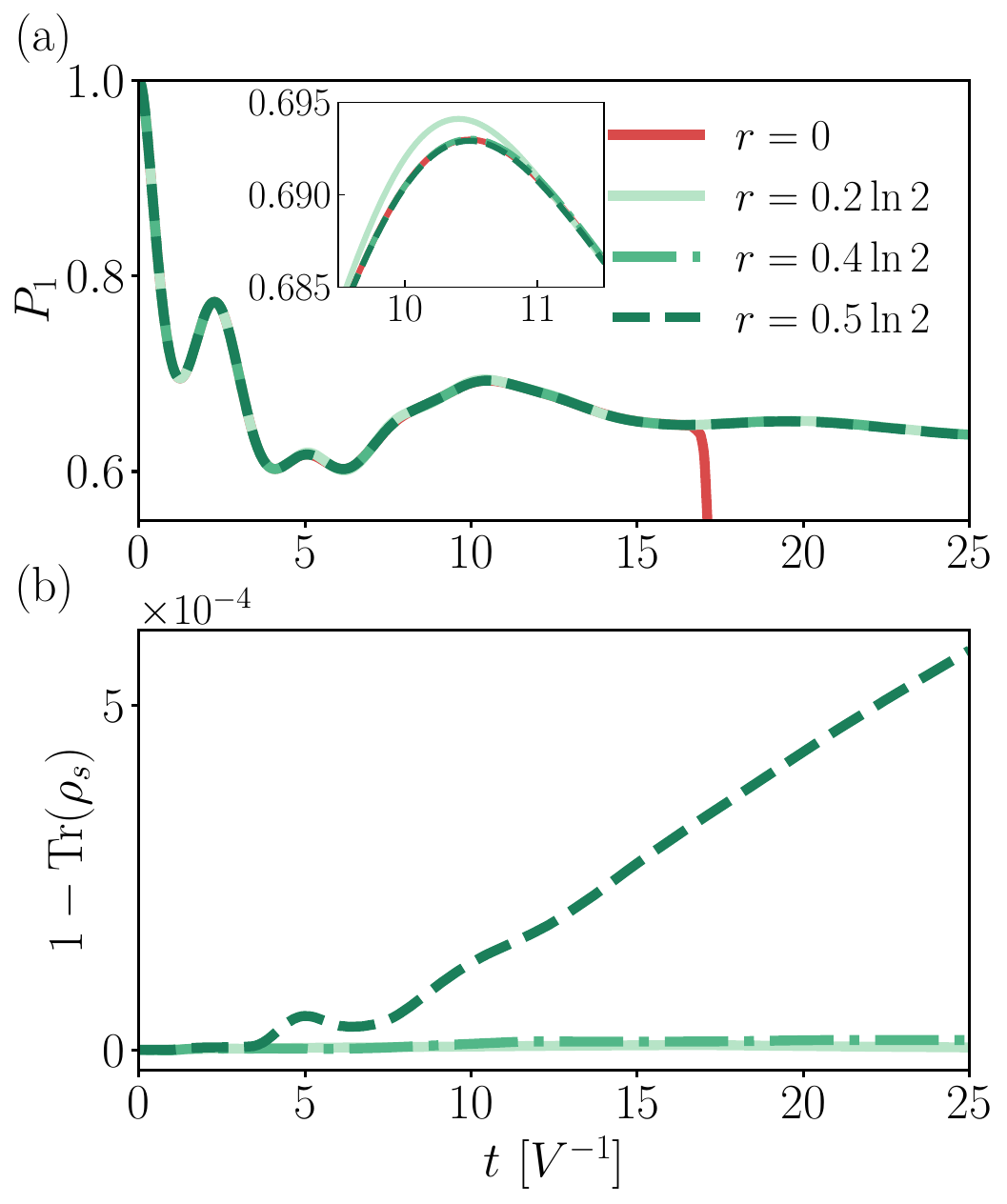}
\caption{
A more stringent test of convergence in our $\mathbf{B}$-$\mathbf{M}^{-1}$-transformed HEOM for the spin--boson model with an underdamped Brownian oscillator spectral density under strong coupling, with $\lambda=3.125$, $\omega_b=0.3$, $\gamma=0.1$, $\beta=1.0$. 
(a) Population of state $|1\rangle$ obtained using conventional HEOM ($r=0$) and $\mathbf{B}$-$\mathbf{M}^{-1}$-transformed HEOM with the $\mathbf{B}$-gauge parameter $r=0.2\ln 2$, $0.4\ln 2$, and $0.5\ln 2$. (b) Trace drift of the reconstructed reduced density operator, $1-\mathrm{Tr}\,\rho_s(t)$, for the three gauge values in panel (a).}
\label{fig5}
\end{figure}

Because atomistic systems display complex spectral densities with both wide and ultranarrow peaks, we extend our analysis to BO spectral densities with finite-width peaks to assess whether the $\mathbf {B}$ transformation also improves numerical stability beyond the undamped-mode limit. The BO spectral density is defined as~\cite{tanimura2012reduced}
\begin{equation}
    J_{\mathrm{BO}}(\omega)=\frac{4\lambda\gamma(\omega_b^2+\gamma^2)\omega}{[(\omega-\omega_b)^2+\gamma^2][(\omega+\omega_b)^2+\gamma^2]}.
\end{equation}
We return to the spin--boson model defined in Eq.~\eqref{eq:sb} and apply the transformation sequence $\mathbf{B}(r)\mathbf{M}^{-1}(\arctan(\omega_b/\gamma))$. Figure~\ref{fig4}(a) shows the low-frequency BO spectral densities at fixed temperature $\beta=V^{-1}$ and reorganization energy $\lambda=3.125V$ and $\omega_b=0.3V$, with $\gamma=0.1V$, $0.2V$, $0.5V$, and $1.0V$. We choose this large reorganization energy ($\lambda=3.125V$) as it embodies a traditionally difficult parameter regime for HEOM. Indeed, as Fig.~\ref{fig4}(b) shows, conventional HEOM diverges for the two narrowest peaks ($\gamma=0.1V$ and $0.2V$). In contrast, the Bogoliubov-transformed HEOM with $r=\frac{1}{2}\ln 2$ remains stable for all four widths over the simulated time interval. This choice recovers the stabilized formulation of Ref.~\onlinecite{li22}, which raises the question of whether interior gauges can improve on it. This behavior is consistent with our expectation that narrow spectral peaks pose greater numerical stability challenges for conventional HEOM. Figure~\ref{fig5} further shows that, among the gauge values tested, $r=0.4\ln 2$ performs best, yielding stable population dynamics while maintaining the trace of the reconstructed reduced density operator closest to unity. Although $r=0.5\ln 2$ yields consistent population dynamics, it exhibits larger trace deviations, whereas $r=0.2\ln 2$ fails to converge, with a root-mean-square error (RMSE) relative to the converged dynamics greater than $10^{-3}$.

For finite-width peaks, one can identify the optimal $\mathbf {B}$ transformation using the low-cost gauge optimization protocol introduced in Ref.~\onlinecite{li2026hidden}. This optimization must ensure that the coefficients multiplying higher-tier ADOs in the reconstruction expressions given in Eq.~\eqref{eq:recons} and the SI have magnitudes less than or equal to $1$.

\textit{Conclusion} --- We have developed a representation-theoretic framework for HEOM based on linear canonical transformations of the auxiliary raising and lowering operators. Combining a generalized Bogoliubov transformation with our previously proposed gauge transformation connects different representations of the same underlying dynamics and encompasses earlier stabilization approaches as special cases. Within a one-parameter subfamily, we have shown that the gauge choice modifies both hierarchy propagation and reduced density operator reconstruction, uncovering a trade-off: increasing the gauge parameter can suppress hierarchy growth while amplifying the contribution of higher-order ADO errors to the reconstructed dynamics. Our analytical and numerical results identify $0<r\leq\frac{1}{2}\ln 2$ as a favorable range for balancing these effects and improving HEOM stability and accuracy.

Recognizing that the Bogoliubov gauge permits a continuous reshaping of HEOM offers a practical advantage over the previously adopted fixed representation at $r=\frac{1}{2}\ln 2$: the gauge can be adjusted to balance hierarchy propagation and reconstruction accuracy. For the undamped single-mode spin--boson model, the transformed hierarchy improves numerical stability and limits the growth of the active hierarchy. For the Holstein model, intermediate gauge choices yield stable population dynamics while reducing the trace drift observed at $r=\frac{1}{2}\ln 2$ by about an order of magnitude. For Brownian oscillator spectral densities with finite-width peaks, the transformed HEOM remains stable over the simulated time interval for all widths examined, including the two narrowest peaks for which conventional HEOM diverges. Here, $r=\frac{1}{2}\ln 2$ recovers the stabilization of Ref.~\onlinecite{li22}, and $r=0.4\ln 2$ retains its stability while reducing the trace drift by more than an order of magnitude, showing that the benefit of continuous gauge selection extends beyond the undamped-mode limit. Thus, recovering an earlier stabilization method as a special case also provides a starting point for improving its numerical performance through continuous gauge selection. Together with our previous work on gauge-optimized HEOM~\cite{li2026hidden}, the present approach expands the available gauge freedom and unifies earlier stabilization methods based on specific phase space representations. This expanded framework provides a broader space in which to search for the optimal HEOM representation for the question being considered.

\section*{acknowledgements}

A.~M.~C., T.~L., and P.~V.~were supported by an Early Career Award in the CPIMS program in the Chemical Sciences, Geosciences, and Biosciences Division of the Office of Basic Energy Sciences of the U.S.~Department of Energy under Award DE-SC0024154. A.M.C.~also acknowledges the support from a David and Lucile Packard Fellowship for Science and Engineering. This work was supported by an Alfred P. Sloan Research Fellowship and the Camille and Henry Dreyfus Foundation through a Camille Dreyfus Teacher-Scholar Award. This work utilized the Alpine high-performance computing resource at the University of Colorado Boulder. Alpine is jointly funded by the University of Colorado Boulder, the University of Colorado Anschutz, Colorado State University, and the National Science Foundation (Award No. 2201538).
\bibliographystyle{plainnat}
\bibliography{quantum}

\pagebreak

\end{document}


\title{Supporting Information for ``Continuous Representation Theory of the Hierarchical Equations of Motion''}

\author{Tianchu Li}
\affiliation{Department of Chemistry, University of Colorado Boulder, Boulder, Colorado 80309, USA\looseness=-1}

\author{Pranay Venkatesh}
\affiliation{Department of Chemistry, University of Colorado Boulder, Boulder, Colorado 80309, USA\looseness=-1}
\author{Andr\'es Montoya-Castillo}
\email{Andres.MontoyaCastillo@colorado.edu}
\affiliation{Department of Chemistry, University of Colorado Boulder, Boulder, Colorado 80309, USA\looseness=-1}
\maketitle
\pagebreak

\section{Gauge-transformed HEOM under general sequences of linear transformations}

We consider two auxiliary modes associated with a single exponential contribution and collect their ladder operators into $\boldsymbol{\Xi}=(\mathbf c^T,(\mathbf c^+)^T)^T$, where $\mathbf c=(c_1,c_2)^T$ and $[c_i,c_j^+]=\delta_{ij}$. The phase-gauge transformation is defined by
\begin{equation}
    \mathcal M(\theta)=
    \begin{pmatrix}
        (M^{-1})^T(\theta)&0\\
        0&M(\theta)
    \end{pmatrix},
    \qquad
    M(\theta)=\frac{1}{\sqrt{2\sin\theta}}
    \begin{pmatrix}
        -e^{-i\theta}&1\\
        -e^{i\theta}&1
    \end{pmatrix},
    \label{eq:si_phase_map}
\end{equation}
where $\sin\theta\neq0$. The Bogoliubov-gauge transformation is
\begin{equation}
    \mathcal B(r)=
    \begin{pmatrix}
        e^{-r}I_2&0\\
        2\sinh r\,I_2&e^rI_2
    \end{pmatrix},
    \label{eq:si_bogoliubov_map}
\end{equation}
where $I_2$ is a $2\times2$ identity matrix. In this work, we restrict our attention to block-lower-triangular transformations. A nonzero upper-right block can generate terms quadratic in the auxiliary raising operators, such as $(c_1^+)^2$, $(c_2^+)^2$, and $c_1^+c_2^+$. These terms introduce additional driving from lower to higher tiers of the hierarchy and can aggravate hierarchy growth and numerical instability.

Any finite sequence of these transformations and their inverses can be written as
\begin{equation}
    \overline{\boldsymbol{\Xi}}=G\boldsymbol{\Xi},\qquad G=\mathcal S_L\cdots\mathcal S_1=\begin{pmatrix}P&0\\Q&P^{-T}\end{pmatrix},\qquad P^TQ=Q^TP,
    \label{eq:si_G_blocks}
\end{equation}
where the rightmost factor acts first. The matrices $P$ and $Q$ encode both the parameters and the order of the transformations. All transposes are ordinary transposes, and the superscript $+$ denotes the canonical creation partner rather than necessarily the Hermitian adjoint. The inverse relations are
\begin{equation}
    \mathbf c=P^{-1}\overline{\mathbf c},\qquad \mathbf c^+=P^T\overline{\mathbf c}^+-P^TQP^{-1}\overline{\mathbf c}.
\end{equation}

We apply this construction independently to each exponential contribution $k$ in complex-pole HEOM~\cite{xu22}, in which the bath correlation function is decomposed as $C(t)=\sum_k d_k e^{-z_k t}$ with amplitudes $d_k$ and exponents $z_k$. We offer the following definitions,
\begin{align}
    Z_k&=\operatorname{diag}(z_k,z_k^*),\qquad A_k=P_k^{-1},\qquad X_k=P_k^TQ_kP_k^{-1},\\
    H_k&=P_kZ_kP_k^{-1},\qquad K_k=X_k^TZ_kP_k^{-1},\qquad a_{j,k}=(A_k)_{1j}+(A_k)_{2j}.
\end{align}
To express the transformed couplings compactly, we introduce the system superoperators
\begin{align}
    \mathcal U_{j,k}Y&=-ia_{j,k}[\sigma_z,Y]+id_k(X_k)_{1j}f(q)Y-id_k^*(X_k)_{2j}Yf(q),\\
    \mathcal V_{j,k}Y&=-id_k(P_k)_{j1}f(q)Y+id_k^*(P_k)_{j2}Yf(q).
\end{align}
The resulting HEOM is
\begin{align}
    \dot{\rho}_{\mathbf m,\mathbf n}&=-\left(i\mathcal L_s+\sum_kH_{11,k}m_k+\sum_kH_{22,k}n_k\right)\rho_{\mathbf m,\mathbf n}\nonumber\\
    &\quad-\sum_k\left(H_{12,k}\sqrt{m_k(n_k+1)}\,\rho_{\mathbf m_k^-,\mathbf n_k^+}+H_{21,k}\sqrt{(m_k+1)n_k}\,\rho_{\mathbf m_k^+,\mathbf n_k^-}\right)\nonumber\\
    &\quad+\sum_k\left(K_{11,k}\sqrt{(m_k+1)(m_k+2)}\,\rho_{\mathbf m_k^{+2},\mathbf n}+K_{22,k}\sqrt{(n_k+1)(n_k+2)}\,\rho_{\mathbf m,\mathbf n_k^{+2}}\right)\nonumber\\
    &\quad+\sum_k(K_{12,k}+K_{21,k})\sqrt{(m_k+1)(n_k+1)}\,\rho_{\mathbf m_k^+,\mathbf n_k^+}\nonumber\\
    &\quad+\sum_k\left(\sqrt{m_k+1}\,\mathcal U_{1,k}\rho_{\mathbf m_k^+,\mathbf n}+\sqrt{n_k+1}\,\mathcal U_{2,k}\rho_{\mathbf m,\mathbf n_k^+}\right)\nonumber\\
    &\quad+\sum_k\left(\sqrt{m_k}\,\mathcal V_{1,k}\rho_{\mathbf m_k^-,\mathbf n}+\sqrt{n_k}\,\mathcal V_{2,k}\rho_{\mathbf m,\mathbf n_k^-}\right).
    \label{eq:general_sequence_heom}
\end{align}
Here, $H_{ij,k}=(H_k)_{ij}$, $K_{ij,k}=(K_k)_{ij}$. Setting $P_k=I_2$ and $Q_k=0$ recovers conventional complex-pole HEOM.

Under a general sequence of these linear transformations, the physical reduced density operator is reconstructed as
\begin{equation}
    \rho_S(t)=\sum_{\mathbf m,\mathbf n}\left[\prod_k w_{m_k,n_k}^{(k)}\right]\rho_{\mathbf m,\mathbf n}(t),
    \label{eq:general_sequence_reconstruction}
\end{equation}
where the weights are determined by the symmetric matrix $W_k=Q_kP_k^{-1}$:
\begin{equation}
    w_{m,n}^{(k)}=\sqrt{m!n!}\,[x^my^n]\exp\left[\frac12(W_k)_{11}x^2+(W_k)_{12}xy+\frac12(W_k)_{22}y^2\right].
    \label{eq:reconstruction_weights}
\end{equation}
Here, $x$ and $y$ are formal variables, and $[x^my^n]$ denotes the coefficient of $x^my^n$ in the Taylor expansion. For the sequence $G_k=\mathcal B(r_k)\mathcal M^{-1}(\theta_k)$, one obtains $W_k=(e^{2r_k}-1)I_2$ and
\begin{equation}
    w_{2u,2v}^{(k)}=\frac{\sqrt{(2u)!(2v)!}}{2^{u+v}u!v!}(e^{2r_k}-1)^{u+v},
\end{equation}
which has zero weight whenever either index is odd. For the sequence $G_k=\mathcal M(\pi/2)\mathcal B(r_k)\mathcal M^{-1}(\pi/2)$ applied in the undamped single-mode case, the reconstruction simplifies to
\begin{equation}
    \rho_S(t)=\sum_{\mathbf n}\left[\prod_k(e^{2r_k}-1)^{n_k}\right]\rho_{\mathbf n,\mathbf n}(t),
\end{equation}
consistent with Eq.~(15) of the main manuscript.

\section{Computational details of HEOM simulations for the Holstein model} 
To improve computational efficiency, we employ a matrix product state (MPS) representation of the HEOM~\cite{shi18, ke22, borrelli21}. We truncate the hierarchy by introducing a local occupation cutoff $N_b$ for each auxiliary mode, retaining only ADOs with $0 \le n_k < N_b$ for all $k$. The matrix elements of the ADOs are represented as 
\begin{align}
    \langle s_b | \rho_{\mathbf n} | s_k \rangle = A_b^{s_b} A_k^{s_k} B^{[0]} B^{[1]} \cdots B^{[NK-1]}, 
\end{align} 
where $s_b$ and $s_k$ label the electronic bra and ket basis states, respectively, and $A_b^{s_b}$ and $A_k^{s_k}$ are the corresponding MPS tensors. Each tensor $B^{[i]}$ represents a bath auxiliary mode, and $NK$ is the total number of auxiliary modes, with $N$ sites and $K$ auxiliary modes per site.

\begin{figure}[htbp]
\renewcommand{\figurename}{FIG.}              
\renewcommand{\thefigure}{S\arabic{figure}}
\centering
\includegraphics[width=14.5cm]{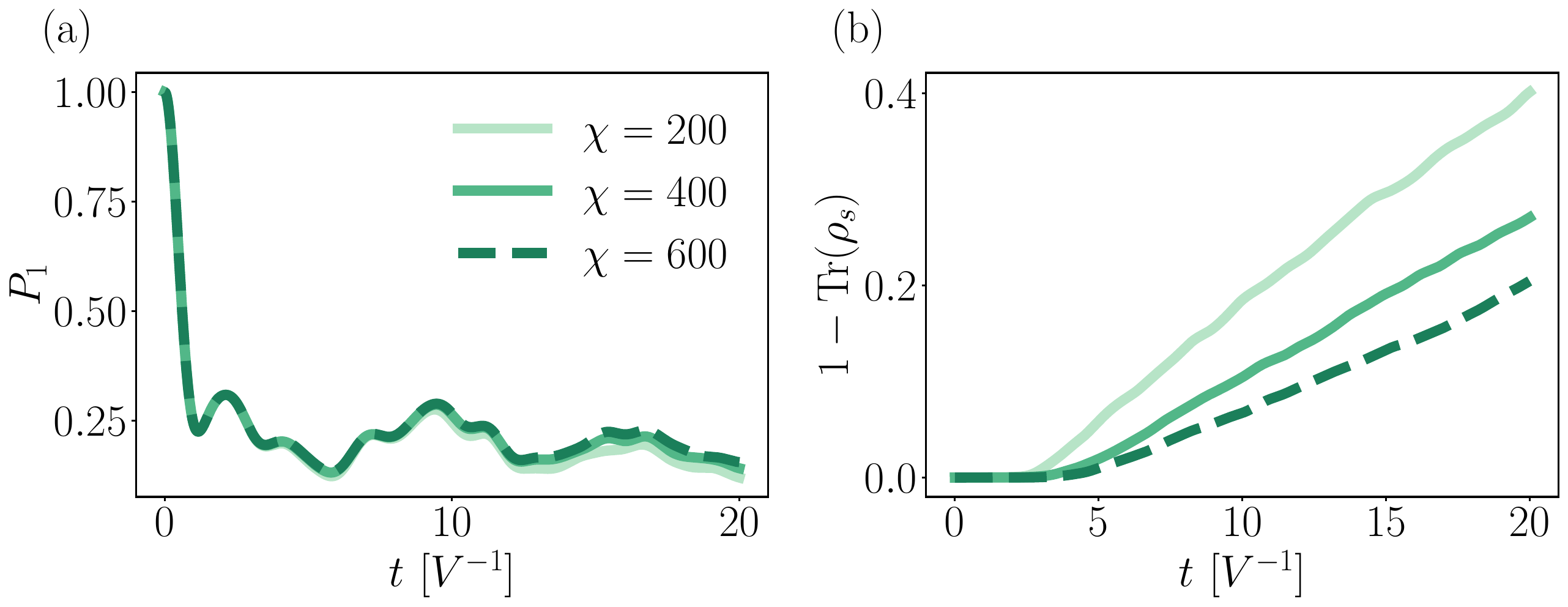}
\vspace{1em}
\caption{(a) Population dynamics of the first site in a five-site Holstein model, calculated using gauge-transformed HEOM with $r=0.5\ln 2$ for different MPS bond dimensions $\chi$. The parameters are $c=\sqrt{2}$, $\omega=1$, and $\beta=1$. (b) Trace drift of the reconstructed reduced density operator, $1 - \mathrm{Tr}(\rho_s(t))$, for the corresponding values of $\chi$.
}
\label{figs1}
\end{figure}

The resulting MPS representation enables efficient propagation of the HEOM using the time-dependent variational principle (TDVP)~\cite{lubich15}. Within this framework, the HEOM equations of motion are projected onto the tangent space of the MPS manifold, yielding time evolution at fixed bond dimensions. All tensor contractions and linear algebra operations are accelerated using GPUs. For all Holstein model simulations, we use a fixed bond dimension of $600$ and a local occupation cutoff of $N_b = 20$. 

Figure~\ref{figs1} compares the population dynamics and the trace of the reconstructed reduced density operator for the Holstein model at different MPS bond dimensions $\chi$, with $r=\frac{1}{2}\ln 2$, corresponding to the stabilization approach of Ref.~\onlinecite{yan20}. Even when the bond dimension is increased to $\chi=600$, the population dynamics remain unconverged, and trace drift persists. These results indicate that increasing the bond dimension alone, within the range considered, is insufficient to obtain converged, trace-preserving reduced dynamics at this gauge choice.

\section{Equilibrium correlation functions in gauge-transformed HEOM}

To obtain equilibrium correlation functions $C(t) =\langle \hat A(t) \hat B(0)\rangle$, we first propagate the imaginary-time HEOM introduced in Ref.~\onlinecite{xing22} to obtain the correlated initial state. In this formulation, the ADOs are denoted by $\rho_{\mathbf k;\mathbf n}$, where $\mathbf k$ labels the imaginary-time auxiliary indices and $\mathbf n$ labels the auxiliary indices of the conventional real-time HEOM. At the end of the imaginary-time propagation, we retain $\rho_{\mathbf 0;\mathbf n}$ and normalize the hierarchy by $\operatorname{Tr}_{\mathrm S}\rho_{\mathbf 0;\mathbf 0}$.

We evaluate these correlation functions using gauge-transformed HEOM by transforming the correlated initial ADOs from the conventional representation to the chosen gauge representation. The two representations correspond to expansions of the same phase-space distribution in different auxiliary bases. The transformation matrix is therefore determined by the overlap between the dual basis of the gauge representation and the basis of the conventional representation, yielding
\begin{equation}
    \widetilde{\rho}_{\mathbf n}^{\mathrm{eq}}=\sum_{\mathbf n'}\mathcal{O}_{\mathbf n,\mathbf n'}(r)\rho_{\mathbf 0;\mathbf n'}^{\mathrm{eq}},\qquad \mathcal{O}_{\mathbf n,\mathbf n'}(r)=\prod_{s=1}^{N}\mathcal{O}_{n_{s,1},n_{s,2};n'_{s,1},n'_{s,2}}(r).
    \label{eq:initial_gauge_overlap}
\end{equation}
For each local auxiliary-mode pair, the overlap matrix elements are
\begin{equation}
    \mathcal{O}_{m,n;m',n'}(r)=\sum_{\ell=0}^{\infty}\delta_{m',m+\ell}\delta_{n',n+\ell}\frac{[-(e^{2r}-1)]^\ell}{\ell!}\sqrt{\frac{(m+\ell)!(n+\ell)!}{m!n!}}e^{-r(m+n+2\ell)}.
\label{eq:initial_gauge_overlap_elements}
\end{equation}
After transforming the equilibrium ADOs, we apply $\hat B$ to each ADO to initialize $\widetilde R_{\mathbf n}(0)=\hat B\widetilde\rho_{\mathbf n}^{\mathrm{eq}}$. We then propagate this hierarchy to time $t$ using the gauge-transformed HEOM and reconstruct the physical operator $R_{\mathrm S}(t)$ using Eq.~(15) in the main manuscript. We recover the equilibrium correlation function by tracing $R_{\mathrm S}(t)$ against $\hat A$: $C_{AB}(t)=\operatorname{Tr}_{\mathrm S}[\hat A R_{\mathrm S}(t)]$.

In this work, we calculate the current autocorrelation function in Holstein model,
\begin{align}
    C_{\rm jj}(t) = \langle \hat j(t) \hat j(0)\rangle ,
\end{align}
where the current is defined as:
\begin{align}
    \hat j=iV\sum_{s=1}^{N}\left(|s+1\rangle\langle s|-|s\rangle\langle s+1|\right).
\end{align}

\section{Holstein dynamics beyond the dilute limit}
We assess the performance of our gauge transformation in fermionic systems beyond the dilute limit on the four-site Holstein model at half filling. The system Hamiltonian is~\cite{stolpp2020charge}
\begin{align}
    H_{\rm sys}=\sum_{m=1}^{N}\left[\epsilon_m n_m-V\left(c_m^\dagger c_{m+1}+\mathrm{H.c.}\right)\right],
\end{align}
where $c_m^\dagger$ and $c_m$ are fermionic creation and annihilation operators satisfying $\{c_m,c_n^\dagger\}=\delta_{mn}$, and $n_m=c_m^\dagger c_m$. We impose periodic boundary conditions. Each site couples to an independent local harmonic oscillator through the coupling operator $f_m=n_m$. At half filling, the two-electron Hilbert space contains $\binom{4}{2}=6$ occupation-number states, with fermionic exchange signs included explicitly in the hopping matrix elements.

\begin{figure}[htbp]
\renewcommand{\figurename}{FIG.}              
\renewcommand{\thefigure}{S\arabic{figure}}
\centering
\includegraphics[width=14.5cm]{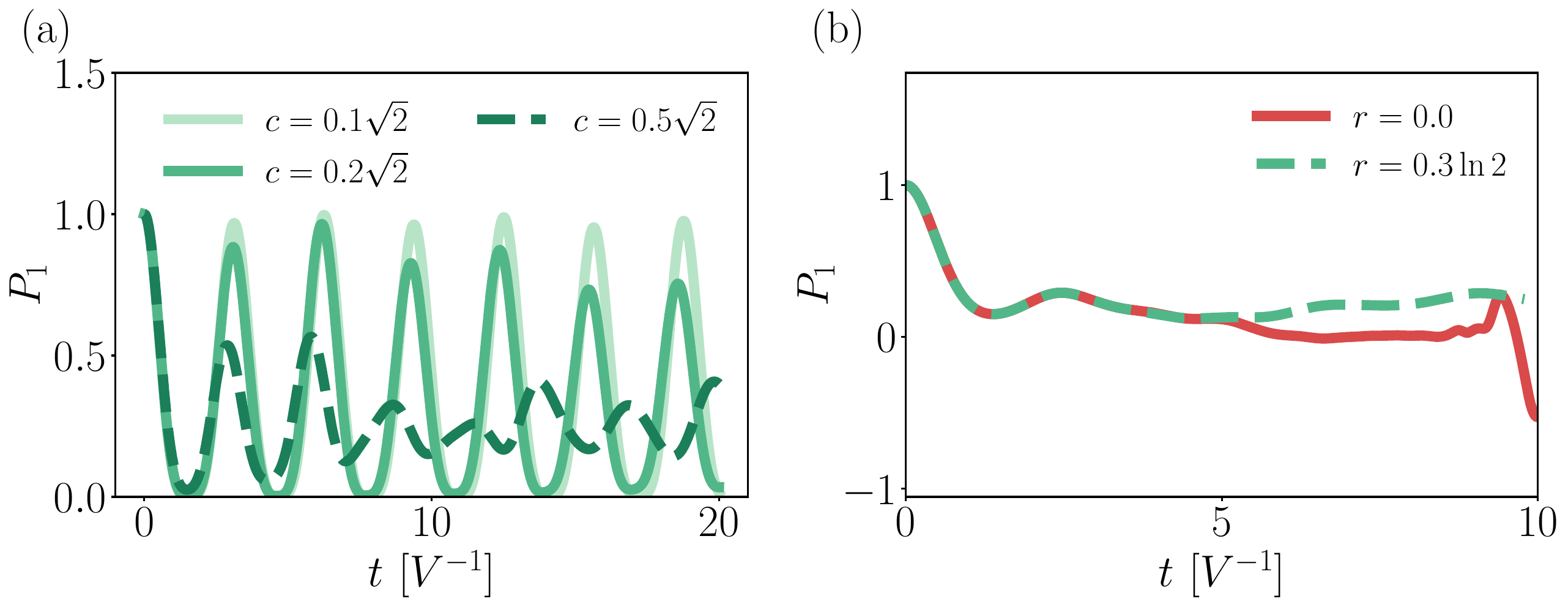}
\vspace{1em}
\caption{Population dynamics of the first site in the Holstein model beyond the dilute limit. (a) Conventional HEOM yields stable dynamics for $c=0.1\sqrt{2}$, $0.2\sqrt{2}$, and $0.5\sqrt{2}$. (b) Comparison of the dynamics obtained using conventional HEOM($r=0$) and gauge-transformed HEOM($r=0.3\ln 2$) at $c=\sqrt{2}$. 
}
\label{figs2}
\end{figure}

Figure~\ref{figs2}(a) shows the Population dynamics of the first site in the Holstein model beyond the dilute limit. Conventional HEOM yields stable dynamics at weaker couplings, $c=0.1\sqrt{2}$, $0.2\sqrt{2}$, and $0.5\sqrt{2}$, but diverges at the stronger coupling $c=\sqrt{2}$. In contrast, gauge-transformed HEOM with $r=0.3\ln 2$ remains stable at this stronger coupling. All calculations use $N_b=10$ and an MPS bond dimension of $\chi=600$. These results demonstrate that gauge-transformed HEOM can stabilize the dynamics of the Holstein model even beyond the dilute limit.

\bibliography{quantum}